\documentclass[pre,twocolumn]{revtex4-1}
\usepackage{amsmath,amssymb}
\usepackage{graphicx}

\begin{document}
\title{Scaling of local roughness distributions}
\author{F. D. A. Aar\~ao Reis\\
Instituto de F\'\i sica, Universidade Federal Fluminense,\\
Avenida Litor\^anea s/n, 24210-340 Niter\'oi RJ, Brazil\\
reis@if.uff.br}
\date{\today}

\begin{abstract}
Local roughness distributions (LRDs) are studied in the growth regimes of lattice models
in the Kardar-Parisi-Zhang (KPZ) class in $1+1$ and $2+1$ dimensions and in a model of
the Villain-Lai-Das Sarma (VLDS) growth class in $2+1$ dimensions.
The squared local roughness $w_2$ is defined as the variance of the height inside a box
of lateral size $r$ and the LRD $P_r\left( w_2\right)$ is sampled as this box glides
along a surface with size $L\gg r$.
The variation coefficient $C$ and the skewness $S$ of the distributions are functions
of the scaled box size $r/\xi\left( t\right)$, where $\xi$ is a correlation length.
For $r \lesssim 0.3\xi\left( t\right)$, plateaus of $C$ and $S$ are observed, but
with a small time dependence.
For a quantitative characterization of the universal LRD, extrapolation of these values
with power-law corrections in time are performed.
The reliability of this procedure is confirmed in $1+1$ dimensions by comparison of
results of the restricted solid-on-solid model and theoretically predicted values of
Edwards-Wilkinson interfaces.
For $r\gg \xi\left( t\right)$, $C$ and $S$ vanish because the LRD converges
to a Dirac delta function.
This confirms the inadequacy of extrapolations of amplitude
ratios to $r\to\infty$, as proposed in recent works.
On the other hand, it highlights the advantage of scaling LRDs by the average instead of
scaling by the variance due to the usually higher accuracy of $C$ compared to $S$.
The scaled LRD of the VLDS model is very close to the KPZ one due to the small difference
between their variation coefficients and the plateaus of $C$ and $S$ are very narrow due
to the slow time increase of $\xi$.
These results suggest that experimental LRDs obtained in short growth times and with
limited resolution may be inconclusive to determine their universality classes
if data accuracy is low and/or data extrapolation to the long time limit is not feasible.
\end{abstract}

%\pacs{81.15.Aa, 05.40.-a, 68.35.Ct, 68.55.-a}
\maketitle

\section{Introduction}
\label{intro}

The study of kinetic roughening helps to understand the basic dynamic mechanisms
of many interface growth processes \cite{barabasi,krug,halpinhealy}, including
important applications to thin film deposition \cite{barabasi,etb}.
A frequent approach is the calculation of scaling exponents of the surface
roughness or of the structure factor and their comparison with the values of
stochastic growth equations.
However, for a more detailed characterization of growing interfaces or for
a study of systems with crossovers, an alternative may be the calculation of
distributions of local and global quantities.
Two decades ago, roughness distributions (RDs) were calculated exactly in
the steady states of linear growth models \cite{foltin,plischke,racz,antalprl,antalpre},
showing connections with other problems \cite{antalprl,bramwell} and experimental
applications \cite{moulinet}.
The RD of the Edwards-Wilkinson (EW) equation \cite{ew} was also calculated as
a function of time in $1+1$ dimensions \cite{antal96}.
Subsequently, numerical works with models in the classes of the Kardar-Parisi-Zhang (KPZ)
equation \cite{kpz} and of the Villain-Lai-Das Sarma (VLDS) equation
\cite{villain,laidassarma} provided accurate steady state RDs in $2+1$ and
higher dimensions \cite{marinari2002,distrib,kelling}.
Amplitude ratios characterizing these RDs typically have small finite-size corrections
\cite{distrib,kelling}.

In all those works, the distributions of the squared global roughness $W_2$
were measured, with $W_2$ defined as the variance of the local height sampled
in the whole interface with a large lateral size $L$.
However, the number of samples in experimental works is usually small, thus it is
difficult to measure a global RD.
For this reason, Antal et al \cite{antalpre} calculated local roughness distributions
in the steady states of one-dimensional Gaussian interfaces, which they called distributions
measured in window boundary conditions (WBC).
The squared local roughness was measured inside a box of size $r$ and the LRD was sampled
as this box glided along a surface with $L\gg r$.
Differences between global RD and LRD were discussed \cite{antalpre}.
Besides the relevance of LRDs to understand the growth kinetics of a given system,
they have other applications; for instance, they are related to the distribution of adhesion
forces of polystyrene microparticles, as discussed in Ref. \protect\cite{you2014}.

On the other hand, thin film or multilayer growth takes place in a very short time regime,
with negligible finite-size effects.
This feature led Paiva and Reis \protect\cite{thereza} to propose the
calculation of LRDs in the KPZ growth regime in $2+1$ dimensions.
Recently, improved estimates of this KPZ LRD were shown to match those of
semiconductor \cite{renanPRB,renanEPL} and organic \cite{hhpal} films grown by
different methods.
Moreover, simulation of the KPZ equation in $1+1$ dimensions showed agreement
with accurate experimental data from turbulent liquid-crystal in five orders of magnitude
\cite{hhtake}.

A subtle point in the comparison of LRDs from models or experiments is the choice of the
range of box size. 
The numerical work of Ref. \protect\cite{thereza} suggested a universal LRD
for several KPZ models in $32\leq r\leq 128$ (measured in lattice units) after growth
times $4000\leq t\leq 8000$ (measured in number of deposition trials).
These values partly guided the choice of box size in the study of semiconductor films
\cite{renanPRB,renanEPL}.
On the other hand, Halpin-Healy and Palasantzas argued that universal LRDs should be
measured under the condition $r\ll\xi$, where $\xi$ is the lateral correlation length
\cite{hhpal}.
This universal LRD could be observed in a wide range of $r$ in $1+1$ and $2+1$ dimensions
after extensive simulation of the KPZ equation that provided very large values of $\xi$
\cite{hhpal,hhtake}.

However, the possible effects of working with limited image resolution, short growth
times, and relatively small correlation lengths were not addressed in previous works on LRDs.
These limitations are frequently present in experimental work, thus the distributions may
have low accuracy or change in time.
For these reasons, the aim of this work is to analyze the time evolution and the box size
dependence of LRDs, with an emphasis on KPZ roughening due to its theoretical and experimental
relevance.
We perform simulations of lattice models in the KPZ class in $1+1$ and $2+1$ dimensions and
a VLDS model in $2+1$ dimensions.
The dimensionless quantities characterizing the relative width and the assymmetry of the
LRDs (variation coefficient and skewness, respectively) depend on the ratio $r/\xi$ beyond
the limit $r\ll \xi$, showing crossovers from the so-called universal LRD values to zero
as $r/\xi$ increases.
Those quantities are expected to form plateaus for small $r/\xi$ (the universal region),
but finite-time corrections may be large and affect their estimates.
This occurs even for growth models that typically show small time and size corrections
in the scaling of the global roughness.
The scaling scenario is confirmed in a VLDS model, with increased limitations to characterize
the universal LRD due to the typically small values of $\xi$.

The rest of this work is organized as follows.
In Sec. \ref{models}, we present the growth models, define basic quantities and
present the main scaling properties of LRDs.
In Sec. \ref{1d}, we present the LRDs for a KPZ lattice model in $1+1$ dimensions
and discuss the consequences of their scaling in time and size.
In Sec. \ref{2d}, this discussion is extended to the KPZ class and to the VLDS class
in $2+1$ dimensions.
Sec. \ref{conclusion} summarizes our results and presents our conclusions. 

\section{Models and basic quantities}
\label{models}

\subsection{Growth equations and lattice models}
\label{eqmodels}

The Kardar-Parisi-Zhang equation \cite{kpz} is
\begin{equation}
{{\partial h}\over{\partial t}} = \nu_2{\nabla}^2 h + \lambda_2
{\left( \nabla h\right) }^2 + \eta (\vec{x},t) ,
\label{kpz}
\end{equation}
where $h(\vec{r},t)$ is a coarse-grained height variable in a
$d$-dimensional substrate, $\nu_2$ is the surface tension, $\lambda_2$ represents the excess
velocity, and $\eta$ is a Gaussian white noise.
The EW equation \cite{ew} is Eq. (\ref{kpz}) with $\lambda_2=0$.

Here we will study two lattice models that are described by the KPZ equation in the
hydrodynamic limit: the restricted solid-on-solid (RSOS) model of Kim and Kosterlitz
\cite{kk} and the etching model of Mello et al \cite{mello}.
They are defined in a hypercubic lattice in $d+1$ dimensions, with a
$d$-dimensional initial flat substrate of lateral size $L$.
The lattice constant is taken as the length unit.
Periodic boundaries are considered in the substrate directions.
Each growth/deposition attempt begins with the random choice of a substrate column $i$.
One time unit corresponds to $L^d$ random selections of columns, independently of the
outcome of the growth attempt.

In the RSOS model, the height of column $i$ increases of one unit if the differences of
heights between nearest neighbor (NN) columns do not exceed $1$.
Otherwise, the growth attempt is rejected.

The etching model is considered in its growth version.
First, the current height $h_0$ of column $i$ is increased by one unit:
$h\left( i\right)\leftarrow h_0+1$.
Subsequently, any NN column whose height is smaller than $h_0$ grows until its
height is $h_0$.

The discrete model in the EW class studied here is the Family model \cite{family}.
In this model, the height of the chosen column $i$ is increased by one unit
only if no NN column has lower height.
If only one NN has a lower height, its height is increased by one unit.
If two or more NN columns have lower heights, one of them is randomly chosen to
grow one unit.

If roughening is dominated by surface diffusion of the adsorbed species,
it is expected to be described by the Villain-Lai-Das Sarma (VLDS)
\cite{villain,laidassarma} equation in the hydrodynamic limit:
\begin{equation}
{{\partial h(\vec{r},t)}\over{\partial t}} = -\nu_4{\nabla}^4 h +
\lambda_{4} {\nabla}^2 {\left( \nabla h\right) }^2 + \eta (\vec{r},t) ,
\label{vlds}
\end{equation}
where $\nu_4$ and $\lambda_{4}$ are constants.

The lattice model in the VLDS class studied here is the conserved RSOS (CRSOS) model
introduced in Ref. \cite{crsosreis}.
All pairs of NN columns obey the condition $|{\Delta h}_{NN}|\leq 1$, but no growth
attempt is rejected in the CRSOS model.
If this condition is satisfied after growth (of one unit) at column $i$, then this
growth attempt is accepted.
Otherwise, a random walk between NN columns is performed until one reaches
a column in which the height increase of one unit satisfies the condition
$|{\Delta h}_{NN}|\leq 1$.
In each step of this walk, the probability to move to any NN column is the same,
independently of the heights of the initial and final columns.
Note that these growth rules are slightly different from those of the original CRSOS
model proposed in Ref. \protect\cite{crsosorig}, but both belong to the same
universality class.

\subsection{Simulation details}
\label{simulation}

% small change
Simulations were performed in lattices with $L=16384$ in $d=1$ and $L=4096$ in $d=2$.
Maximal times are of order ${10}^4$ for all models.
For the RSOS and Family models in $d=1$, ${10}^5$ and ${10}^3$ different deposits
were respectively grown.
For the etching and RSOS models in $d=2$, $100$ different deposits were grown,
and $10$ deposits were grown for the CRSOS model.

LRDs are calculated at $13$ selected times in square boxes of sizes from $16$ to $1024$
in $d=2$ (linear boxes in $d=1$).
For each configuration of the deposit and each value of $r$, the gliding box
is allowed to occupy all possible positions over the surface; thus, each column will
be placed inside the box at $r^d$ different positions.
At each position of the box, the squared roughness $w_2$ is defined as the variance
of the heights of the columns inside it.
$P_r(w_2)dw_2$ is the probability that the squared roughness $w_2$ is in the
interval $[w_2,w_2+dw_2]$.   

\subsection{Correlation length}
\label{correlation}

For calculating the lateral correlation length, we begin by calculating the
autocorrelation function
\begin{equation}
\Gamma\left( s,t\right) \equiv \langle {\left[ \tilde{h}\left( {\vec{r}}_0+\vec{s},t\right)
\tilde{h}\left( {\vec{r}}_0,t\right) \right]}^2 \rangle \qquad ,\qquad s\equiv |\vec{s}| ,
\label{defcorr}
\end{equation}
where $\tilde{h}\equiv h-\overline{h}$.
The configurational averages are taken over different initial positions ${\vec{r}}_0$,
different orientations of $\vec{s}$ (substrate directions), and different deposits.

In the CRSOS model, mounded surface structure is observed, thus one may estimate the
correlation length $\xi\left( t\right)$ as the first zero of
$\Gamma\left( s,t\right)$ \cite{renanPRB,siniscalco}.
We will refer to this length as $\xi_0\left( t\right)$.
From $t=100$ to $t=4\times{10}^4$, $\xi_0$ varies from $12$ to $76$, which is
consistent with the small dynamical exponent $z\approx 3.3$ \cite{laidassarma,crsosreis}.

In the KPZ and RSOS models, $\Gamma\left( s,t\right)$ frequently oscillates
with $s$ before crossing the value $\Gamma =0$.
Thus we use $\Gamma\left( \xi_1,t\right) /\Gamma\left( 0,t\right) = 0.1$ for
estimating the correlation length $\xi_1\left( t\right)$.
For the RSOS model in $d=1$, $\xi_1$ varies from $23$ to $557$ as $t$ increases
from $100$ to $12800$, which is consistent with the dynamical exponent $z=1.5$ \cite{kpz};
in $d=2$, $\xi_1$ varies from $12$ to $230$ in the same time range, which is
consistent with $z\approx 1.63$ \cite{kelling,kpz2d}.

\begin{figure}[!h]
\includegraphics[width=7.5cm]{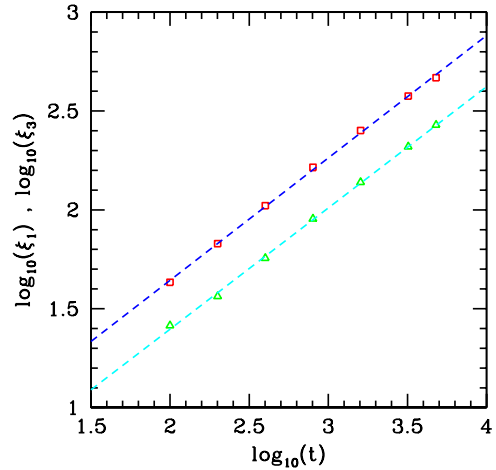}
\caption{(Color online) Correlation lengths $\xi_1$ (squares) and $\xi_3$ (triangles)
as a function of time for the etching model in $2+1$ dimensions. Dashed lines are
least squares fits of each set of data points, respectively with slopes $0.618$
and $0.613$.
}
\label{figxi}
\end{figure}

The most suitable method to estimate the lateral correlation length in these growth
models actually depends on details of the interface morphology \cite{etb,siniscalco}.
The reliability of our approach is tested with the alternative use of a correlation
length $\xi_3\left( t\right)$ defined by
$\Gamma\left( \xi_3,t\right) /\Gamma\left( 0,t\right) = 0.3$.
Fig. \ref{figxi} shows the time evolutions of $\xi_1$ and $\xi_3$ for the etching model
in $d=2$, which confirms that both lengths scale as $t^{1/z}$ with the expected dynamical
exponent.
$\xi_3$ is $30\%$ to $50\%$ smaller than $\xi_1$ in the models studied here.

\section{Local roughness distributions in $1+1$ dimensions}
\label{1d}

\subsection{Simulation results for the RSOS model}
\label{simulation1d}

The expected scaling of RDs \cite{foltin,racz} can be extended to LRDs as
\begin{equation}
P_r\left( w_2\right) = \frac{1}{\langle w_2\rangle}
\Psi\left( \frac{w_2}{\langle w_2\rangle} \right) .
\label{scalingaverage}
\end{equation}

\begin{figure}[!h]
\includegraphics[width=8.5cm]{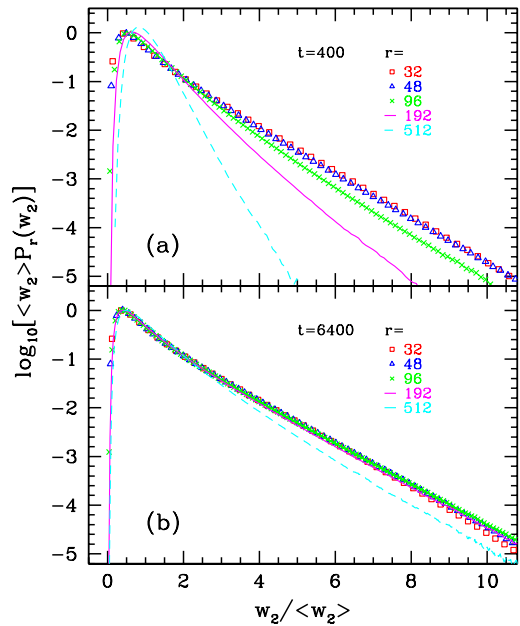}
\caption{(Color online) Scaled LRDs of the RSOS model in $d=1$ for the listed box
sizes at (a) $t=400$ and (b) $t=6400$.
}
\label{RDrsos1d}
\end{figure}

% changes
Figs. \ref{RDrsos1d}a and \ref{RDrsos1d}b shows LRDs of the RSOS model in five
box sizes at $t=400$ and $t=6400$, respectively, scaled according Eq. (\ref{scalingaverage}).
Those plots show that scaled LRDs collapse in certain ranges of $r$.
The range begins in $r\sim 30$ and ends at a value that depends on time.
At $t=400$ (Fig. \ref{RDrsos1d}a), the collapse of curves with $r=32$ and $r=48$ is
observed, but there are large discrepancies with the curves of larger $r$.
At $t=6400$ (Fig. \ref{RDrsos1d}b), good data collapse in two orders of $P\left( w_2\right)$
is observed up to $r=192$, with small deviations appearing for smaller $P\left( w_2\right)$.
A significant deviation is observed only for $r=512$.

When LRDs measured in different growth times are compared, differences between the apparently
universal curves of short and long times are also observed.
This is observed in Figs. \ref{RDrsos1d}a,b, which have the same ranges in both axis:
the tail of the collapsed curves at $t=400$ ($r=32$ and $48$) is slightly steeper than
the tail of the collapsed curves at $t=6400$ ($r=32$, $48$, and $96$).
In linear-linear plots (not shown), it is observed that the scaled LRDs at $t=400$ also
have smaller peaks.
Thus, the collapse observed in Fig. \ref{RDrsos1d}a for small $r$ and short time is not
suitable for a quantitative characterization of the universal LRD, although it may be viewed
as an anticipation of the existence of a universal distribution (as proposed in Ref.
\protect\cite{thereza} for $d=2$).
On the other hand, for $t=6400$ and $t=25600$, good collapse in five orders of magnitude
of $P\left( w_2\right)$ is observed, indicating a convergence to the universal LRD.

As $r$ increases, Figs. \ref{RDrsos1d}a,b shows that the scaled LRD becomes narrower.
This corresponds to a decrease in the variation coefficient
\begin{equation}
C \equiv \frac{\sigma}{\langle w_2\rangle} ,
\label{defk}
\end{equation}
where
\begin{equation}
\sigma \equiv {\left( {\langle {w_2}^2\rangle} - {\langle w_2\rangle}^2 \right) }^{1/2} .
\label{sigma}
\end{equation}

\begin{figure}[!h]
\includegraphics[width=8.5cm]{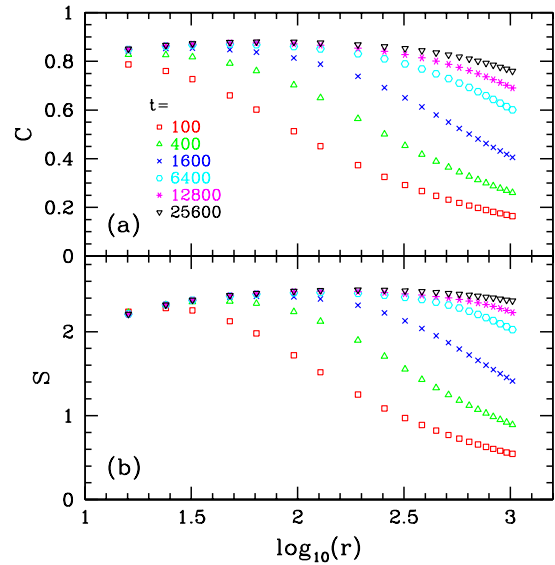}
\caption{(Color online) (a) Variation coefficient and (b) skewness of the LRDs of
the RSOS model in $d=1$ as a function of box size at the listed times.
}
\label{ksrsos1d}
\end{figure}

Fig. \ref{ksrsos1d}a shows $C$ as a function of $r$ for several times.
At the longest times, it shows the formation of plateaus of $C$ in certain ranges of $r$,
which is characteristic of a universal (box size independent) LRD.
For $t=400$, the two curves that collapsed in Fig. \ref{RDrsos1d}a had $C$ between $0.75$
and $0.81$, which are significantly smaller than the values of the long-time plateaus.
For $t=1600$, the range $16\leq r\leq 64$ gives a plateau with $C=0.84\pm 0.01$.
Results for longer times show wider plateaus with slightly larger values of $C$.
Careful inspection of Fig. \ref{ksrsos1d}a shows that finite-time corrections appear even
when data for $t=12800$ and $t=25600$ are compared.
For $t=25600$, $C=0.87\pm 0.01$ is obtained in more than one decade or $r$.

The LRDs also get a more symmetric shape as $r$ increases.
This is related to a decrease of the skewness of the distribution, which is defined as
\begin{equation}
S \equiv \frac{\langle{\left( w_2 - \langle w_2\rangle \right) }^3\rangle}
{{\langle{\left( w_2 - \langle w_2\rangle \right) }^2\rangle}^{3/2}} . 
\label{skewness}
\end{equation}
Fig. \ref{ksrsos1d}b shows $S$ as a function of $r$ for several times.
The presence of plateaus is also observed in those plots, but they are smaller and shifted
to larger values of $r$ when compared to the plateaus of $C$ (Fig. \ref{ksrsos1d}a).
This feature reduces the range of $r$ with constant $C$ and $S$, which is the range of the
universal LRD.
The values of $S$ in the plateaus also show a non-negligible time dependence:
for $t=1600$, an estimate $S=2.36\pm 0.06$ for the universal LRD is suggested;
however, for $t=25600$, the plateau formed at larger box sizes gives $S=2.49\pm 0.01$.

\begin{figure}[!h]
\includegraphics[width=8.5cm]{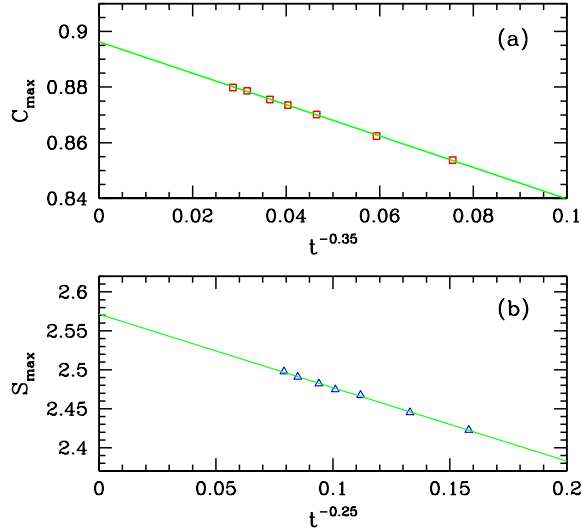}
\caption{(Color online) Extrapolation in time of the maximal variation coefficient (a)
and skewness (b) of the LRDs of the RSOS model in $d=1$. The variables in the abscissa
have the exponents that give the best linear fits (solid lines) of each data set.
}
\label{extrapcsrsos1d}
\end{figure}

For each time, maximal values $C_{max}\left( t\right)$ and $S_{max}\left( t\right)$
were obtained, typically in the middle of the above mentioned plateaus.
They were plotted as a function of $t^{-\lambda_C}$ and $t^{-\lambda_S}$,
respectively, for several values of exponents $\lambda_C$ and $\lambda_S$.
For $t\geq 1600$, the exponents that provided the best linear fits of each set of data
were $\lambda_C=0.35$ and $\lambda_S=0.25$, as shown in Figs. \ref{extrapcsrsos1d}a,b.
Those fits provide extrapolated estimates ($t\to\infty$) $C_{max}=0.896\pm 0.05$ and
$S_{max}=2.57\pm 0.02$.

The fits with small exponents $\lambda_C$ and $\lambda_S$ are clear indications of large
scaling corrections.
Estimates of $C$ and $S$ at $t=400$ are almost $10\%$ smaller than the asymptotic ones.
Since the RSOS model has relatively weak corrections in the roughness scaling, these results
suggest that other models and experimental data may also show large scaling corrections in
their LRDs.
A comparison of these estimates with analytical results for Gaussian interfaces and 
with recent numerical data for the KPZ equation is postponed to Sec. \ref{comparison}.

\subsection{Scaling in the box size}
\label{scaling}

Kinetic roughening theory predicts that the height fluctuations of an interface are
highly correlated in distances $l\ll\xi$, but uncorrelated at distances $l\gg \xi$.
The exact solution of the EW and other linear stochastic equations illustrate
this feature \cite{ew,krug}.

% small change
For this reason, we expect that the scaled LRDs in boxes of sizes
$r\ll\xi$ are universal and represent these highly correlated regions.
This condition was proposed in Ref. \protect\cite{hhpal} and was used for choosing LRDs
for the comparisons with experimental data in $d=1$ and $d=2$ \cite{hhpal,hhtake}.
It parallels the universality of the scaled steady state RDs in systems of
different lateral sizes $L$, since all wavelengths (up to $\sim L$) are excited
in that state \cite{foltin,racz,antalpre}.
An additional condition to observe this feature is $r\gg 1$ because this
is necessary for a hydrodynamic description to be valid.

On the other hand, when $r\gg \xi\left( t\right)$, a single box is probing a region
in which most local heights are uncorrelated.
For KPZ in $d=1$, these heights fit to a Tracy-Widom distribution enhanced by a
factor that depends on time (as $t^{1/3}$) and on the model, as predicted in
Refs. \protect\cite{sasamoto,calabrese} and confirmed numerically in Refs.
\protect\cite{tiagokpz1d2013,hhpre2014}.
The width of this height distribution is the global roughness at time $t$,
$W\left( t\right)$, which has a well defined value for a given KPZ model at a given time.
The corresponding LRD is non-zero at a single value $w_2={W\left( t\right)}^2$,
thus the scaled LRD is $\Psi\left( x\right)=\delta\left( x-1\right)$. 

\begin{figure}[!h]
\includegraphics[width=8.5cm]{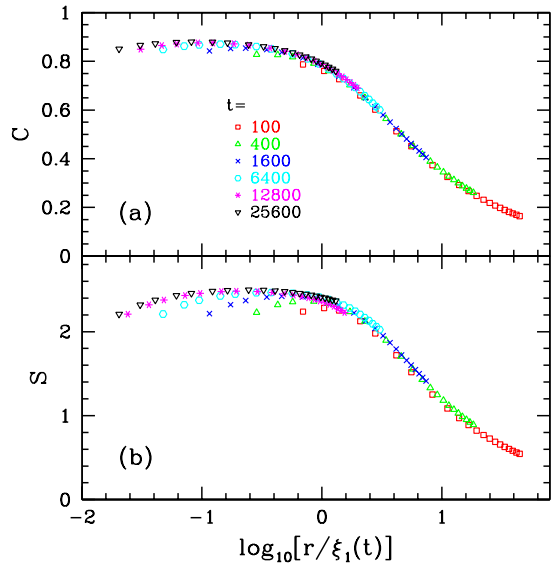}
\caption{(Color online) (a) Variation coefficient and (b) skewness of the LRDs of
the RSOS model in $d=1$ as a function of the scaled box size at the listed times.
}
\label{ksscaledrsos1d}
\end{figure}

This reasoning suggests that that the amplitude ratios which characterize the LRD
scale with $r/\xi$.
Figs. \ref{ksscaledrsos1d}a and \ref{ksscaledrsos1d}b show $C$ and $S$, respectively,
as a function of $r/\xi_1\left( t\right)$.
The scaling is confirmed by the good data collapse, with a universal crossover between
the stationary LRD and the delta distribution.
The time evolution of the maximal values of $C$ and $S$ highlighted in
Sec. \ref{simulation1d} (Figs. \ref{extrapcsrsos1d}a,b) occurs while the plateaus advance
to smaller $r/\xi_1$ (left side of Figs. \ref{ksscaledrsos1d}a,b).
Those plateaus are observed for $r/\xi_1 \lesssim 0.3$
[$\log{\left( r/\xi_1\right)}\leq -0.5$], which is a condition less restrictive than
requiring $r\ll\xi$.
This is consistent with the time-increasing range of $r$ in which the universal LRD is found.

A good data collapse is also observed when $C$ and $S$ are plotted as a function of
$r/\xi_3$ (not shown here).
This is expected because $\xi_1$ and $\xi_3$ scale in time with the same exponent
(Sec. \ref{correlation}).
The only change is a small displacement in the range of $r/\xi$ in which the
plateaus of $C$ and $S$ are observed.

The convergence of $C$ and $S$ to zero is not shown in Figs. \ref{ksscaledrsos1d}a,b
because much larger box sizes $r$ would be necessary, requiring simulations
in much larger lattices.

\subsection{Comparison with steady state and EW distributions}
\label{comparison}

The steady state of the KPZ class in $d=1$ has the same global RD of the EW class
\cite{foltin}.
The distribution in periodic boundaries has generating function
$G_{PB}\left( x\right)=\sqrt{6x}/\sinh{\sqrt{6x}}$ \cite{foltin,antalpre},
with $x\equiv w_2/\langle w_2\rangle$, which
gives $C=\sqrt{2/5}\approx 0.632$ and $S=4\sqrt{10}/7\approx 1.807$.
A more relevant comparison is that with distributions of the local roughness in WBC
at the steady state by Antal et al \cite{antalpre}, which are obtained in boxes of size $r\ll L$.
The generating function is
$G_{WBC}\left( x\right)=\sqrt{\sqrt{12x}/\sinh{\sqrt{12x}}}$ \cite{antalpre},
which gives $C=2/\sqrt{5}\approx 0.894$ and $S=8\sqrt{5}/7\approx 2.556$.
This distribution differs from the global one.
We confirmed these estimates by performing simulations of the RSOS model in the steady
state.

Ref. \protect\cite{hhtake} showed the data collapse of KPZ LRDs in the growth
regime and the distrubution of Antal et al in WBC.
The agreement was also supported by the estimate of the skewness $2.55$ in a plateau in
the range of box size $16\leq r\leq 96$.
These results suggest that the KPZ nonlinearity does not affect the universal LRDs in
the growth regime.
Estimates of $C$ were not presented in that work; estimates of the kurtosis varied with $r$,
but the accuracy of higher moments of numerically calculated distributions are actually
expected to be lower. 

Our long time (extrapolated) estimates of $C_{max}$ and $S_{max}$ also agree with the
values of Antal et al in the steady state WBC \cite{antalpre},
confirming the proposal of Ref. \protect\cite{hhtake}.
However, as shown in Sec. \ref{simulation1d}, these estimates are obtained in the
RSOS model after accounting for large scaling corrections; instead, the plateaus of $C$ and
$S$ at the longest simulated time ($t=25600$) are differ from the values of Antal et al
\cite{antalpre}.

\begin{figure}[!h]
\includegraphics[width=8.5cm]{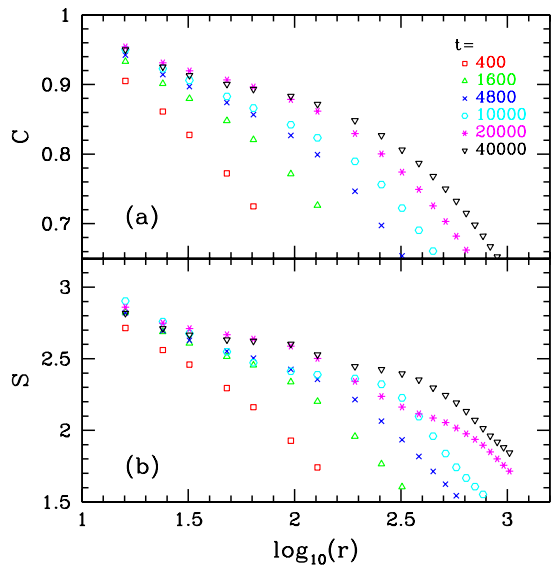}
\caption{(Color online) (a) Variation coefficient and (b) skewness of the LRDs of
the Family model in $d=1$ as a function of the box size at the listed times.
}
\label{ksfam1d}
\end{figure}

The Family model is in the EW class and also has small corrections in the scaling of
the average roughness.
However, the convergence of the LRDs of this model is even slower than
that of the RSOS model.
Figs. \ref{ksfam1d}a,b show $C$ and $S$ for the LRD of the Family model.
Until $t\sim {10}^4$, both quantities are monotonically decreasing with $r$.
At $t={10}^4$, the linear fit of the $C$ and $S$ data with $32\leq r\leq 128$ 
has a very small slope; at $t= 2\times {10}^4$, the slope of this fit is even smaller;
this evolution suggests that
plateaus will be formed at longer times with $0.88\geq C\geq 0.92$ and $2.5\leq S\leq 2.7$.
These values are consistent with the steady state ones of Antal et al \cite{antalpre}.
However, a systematic extrapolation to $t\to\infty$ cannot be performed with this model
data, thus these estimates may be viewed only as a guess based on the apparent trends
at long times.

The global RD of an EW interface calculated in Ref. \protect\cite{antal96} has vanishing
width ($C\to0$) at times $t$ much smaller than the time of relaxation to the steady state.
This is the case of the growth regime studied here.
It is consistent with our results because the global RD is measured in
$r=L\gg \xi$, supporting the extension of our  scaling arguments to other growth classes.

\subsection{Distributions scaled by the variance}
\label{alternative}

When steady state RDs are calculated, they are frequently scaled by the variance
according to the relation
\begin{equation}
P\left( w_2\right) = \frac{1}{\sigma}
\Phi\left( \frac{w_2-\langle w_2\rangle}{\sigma} \right) .
\label{scalingvariance}
\end{equation}
The function $\Phi$ is advantageous over $\Psi$ [Eq. (\ref{scalingaverage})]
for reducing finite-size effects in numerical data \cite{intrinsic}.
For this reason, scaling by the variance was formerly used with LRDs
\cite{thereza,renanPRB,renanEPL,hhpal,hhtake}.

However, the function $\Phi$ has unit variance, while the variance of the function
$\Psi$ is $C^2$.
Thus, the crossover features of the relative width $C$ are hidden by
Eq. (\ref{scalingvariance}).
Since estimates of $C$ are typically more accurate than those of $S$ or of the kurtosis,
relevant information may be lost in scaling by the variance.

\begin{figure}[!h]
\includegraphics[width=8.5cm]{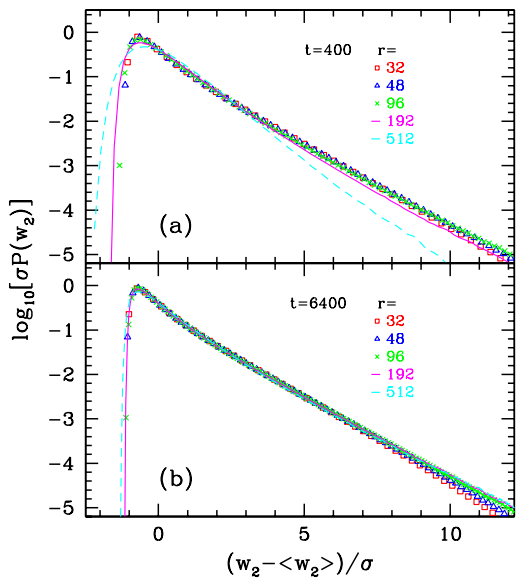}
\caption{(Color online) LRDs of the RSOS model in $d=1$ scaled by the variance
at (a) t=400 and (b) t=6400, for the same box sizes of of Figs. \ref{RDrsos1d}a,b.
}
\label{RDsigmarsos1d}
\end{figure}

This is confirmed in Figs. \ref{RDsigmarsos1d}a,b, which show the same LRDs of Figs.
\ref{RDrsos1d}a,b scaled by the variance.
For $t=400$, deviations from the universal LRD for large $r$ are less striking
in Fig. \ref{RDsigmarsos1d}a when compared to Fig. \ref{RDrsos1d}a.
For $t=6400$, the peaks and the right tails for $r\leq 512$ show good collapse in
scaling by the variance (Figs. \ref{RDsigmarsos1d}b); instead, scaling by the average
(Fig. \ref{RDrsos1d}b) highlights the deviations for the largest $r$.

With scaling by the variance, the differences between the scaled distributions
are quantitatively measured by the skewness $S$ or by dimensionless ratios of
higher order moments, such as the kurtosis calculated
in Refs. \protect\cite{thereza,renanPRB,renanEPL,hhpal,hhtake}.
However, their accuracy is usually smaller than that of the variation coefficient $C$.
Note that this is a particular feature of the LRDs in the growth regime because there
is a crossover in the shape of the distribution at $r/\xi \sim 1$.

\section{Local roughness distributions in $2+1$ dimensions}
\label{2d}

\subsection{KPZ class}
\label{kpzsection}

The collapse of LRDs of the etching and RSOS models was illustrated in
Ref. \protect\cite{thereza}, which was the basis to argue that there is a universal
distribution of the KPZ class.
In that work, the skewness $S$ and kurtosis $Q$ of the LRDs were also extrapolated
to $r\to\infty$ in order to charatectize them in the hydrodynamic limit.
Similar procedure was adopted in Refs. \protect\cite{renanPRB,renanEPL}.

However, Ref. \protect\cite{hhpal} argues that this extrapolation is not reliable
since the condition $r\ll \xi$ fails for large $r$. 
This is consistent with the discussion of Sec. \ref{scaling}, which shows that
$C$ and $S$ decrease to zero for very large $r$ (the same occurs with $Q$).
For this reason, here we emphasize the scaling of $C$ and $S$ on $r/\xi$ to
determine the universal (stationary; small $r$) LRD and its crossover features.

\begin{figure}[!h]
\includegraphics[width=8.5cm]{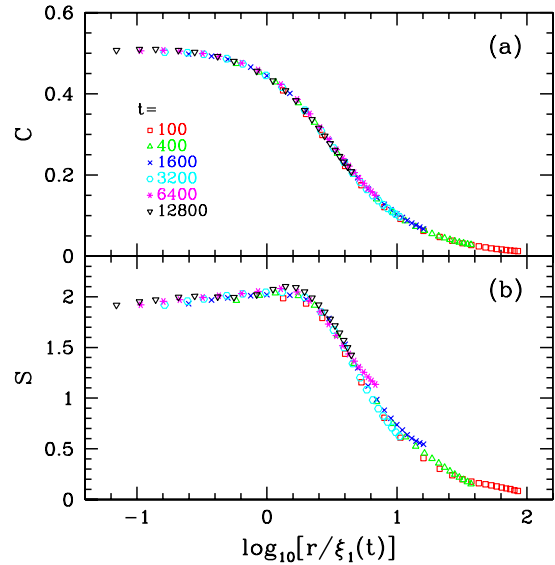}
\caption{(Color online) (a) Variation coefficient and (b) skewness of the LRDs of
the RSOS model in $d=2$ as a function of the scaled box size at the listed times.
}
\label{ksscaledrsos2d}
\end{figure}

Figs. \ref{ksscaledrsos2d}a,b show $C$ and $S$ of the
LRDs of the RSOS model as a function of the scaled box size $r/\xi_1$.
The data collapse confirms the scaling proposed in Sec. \ref{scaling}, although a tiny
time-dependence of the height of the plateaus indicates the presence of scaling corretions.
The decrease of $C$ and $S$ for large $r/\xi_1$ confirms the convergence of the
LRD to a Dirac delta function.
The plateaus of $C$ and $S$ in Figs. \ref{ksscaledrsos2d}a,b for the longest time
($t=12800$) give estimates $0.49\leq C\leq 0.51$ and $1.97\leq S\leq 2.10$ for the
universal LRD.

\begin{figure}[!h]
\includegraphics[width=8.5cm]{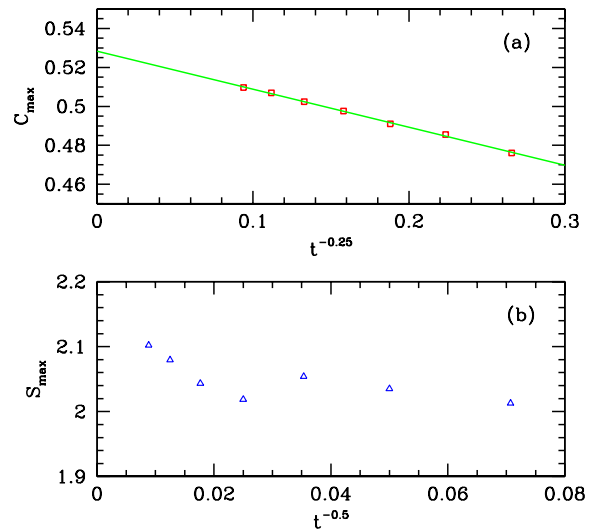}
\caption{(Color online) Extrapolation in time of the maximal variation coefficient (a)
and skewness (b) of the LRDs of the RSOS model in $d=2$. The variable in the abscissa
in (a) has the exponent that gives the best linear fit (solid line) of the data.
}
\label{extrapcsrsos2d}
\end{figure}

However, accounting for finite-time corrections is essential, similarly to the
one-dimensional model.
The maximal values $C_{max}\left( t\right)$ and $S_{max}\left( t\right)$ are obtained
for several times and plotted as a function of $t^{-\lambda_C}$ and $t^{-\lambda_S}$
for several values of these exponents.
In Fig. \ref{extrapcsrsos2d}a, we show the extrapolation of $C_{max}$ with
$\lambda_C=0.25$, which provides the best linear fit of the data for $t\geq 200$.
Our asymptotic estimate is $C=0.53\pm 0.02$, which is larger than that predicted in the
longest simulated time.
On the other hand, Fig. \ref{extrapcsrsos2d}b shows a non-monotonic evolution of
$S_{max}$.
The exponent $\lambda_S =0.5$ was used because it fits the data for the longest times,
but it is difficult to know wether $S$ will converge with this scaling corrections
or will oscillate at longer times.
In any case, the evolution of the data in Fig. \ref{extrapcsrsos2d}b suggests that
the asymptotic $S$ is between $2.0$ and $2.2$.

The estimate of $S$ in Ref. \protect\cite{thereza} is below this range due
to finite-time corrections that were not considered in that work.
The estimate $S=2.03$ of Refs. \protect\cite{hhpal,hhtake} is included in that range.
The study of the time evolution of $C$ seems to improve the comparison of
LRDs due to its smaller statistical fluctuations.

\begin{figure}[!h]
\includegraphics[width=8.5cm]{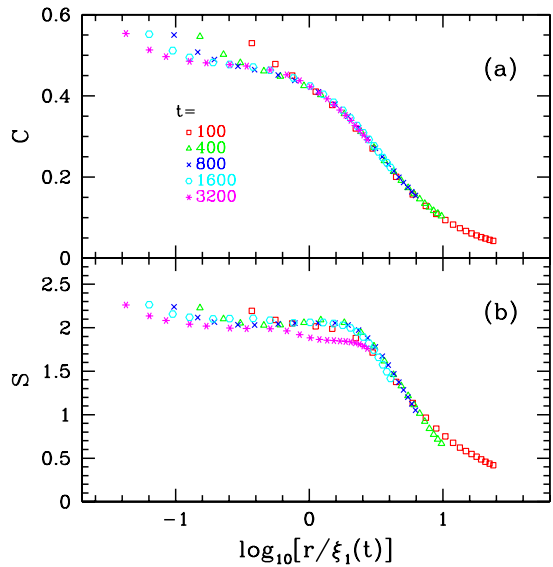}
\caption{(Color online) (a) Variation coefficient and (b) skewness of the LRDs of
the etching model in $d=2$ as a function of the scaled box size at the listed times.
}
\label{ksscaledetch2d}
\end{figure}

Figs. \ref{ksscaledetch2d}a,b show $C$ and $S$ of the LRDs of the etching model.
The data is limited to $t\leq 3200$ because flucutations are very large at longer times.
In the smallest boxes ($r\leq 32$), deviations from the plateaus of $C$ and $S$
are observed.
Those plateaus have larger fluctuations than those of the RSOS model
(Figs. \ref{ksscaledrsos2d}a,b), but they are consistent with the ranges
of $C$ and $S$ presented above.

The etching model also has small corrections in the global roughness scaling
when compared to most KPZ models \cite{kpz2d}.
Thus, the large fluctuations and significant time-dependence shown in Figs.
\ref{ksscaledetch2d}a,b suggest that comparison of LRDs of more complex models
or of thin film surfaces may also have large scaling corrections.

A direct comparison of scaled data for the RSOS and the etching models is not shown
due to the arbitrariness in the calculation of the correlation length $\xi_1$.
The collapsed data in Figs. \ref{ksscaledetch2d}a,b are slightly shifted to the
right in comparison with the data in Figs. \ref{ksscaledrsos2d}a,b.

The above results also show that the condition $r\ll \xi$ proposed in Ref.
\protect\cite{hhpal} may be very restrictive.
The plateaus of $C$ and $S$ of the RSOS model are observed for
$\log_{10}{\left( r/\xi_1\right)}\lesssim 0.5$, i. e. $r\lesssim 0.3\xi_1$,
similarly to the case $d=1$.
This may be particularly important for working with microscopy images in which the
conditions $r\gg 1$ (in number of pixels) and $r\ll \xi$ cannot be simulteneously
fulfilled.
However, the extrapolations of $C_{max}$ and $S_{max}$ are essential to compare
those systems LRDs with theoretical ones.

\subsection{VLDS class}
\label{vldssection}

\begin{figure}[!h]
\includegraphics[width=8.5cm]{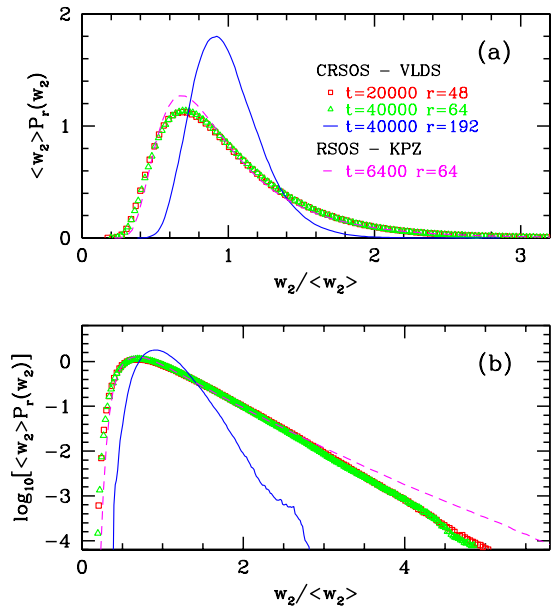}
\caption{(Color online) Scaled LRDs of the CRSOS model in $d=2$ in (a) linear-linear
and (b) log-linear scale at the listed times and box sizes.
A scaled LRD of the RSOS model in $d=2$ is also shown for comparison.
}
\label{RDcrsos2dgeral}
\end{figure}

The correlation length of the CRSOS model increases in time very slowly because
the exponent $z$ is large, as shown in Sec. \ref{models}.
Thus, a collapse of LRDs is obtained only for small $r$ even at
very long times ($t>{10}^4$).
This is illustrated in Figs. \ref{RDcrsos2dgeral}a and \ref{RDcrsos2dgeral}b, with
LDRs plotted in linear-linear and log-linear scales, respectively:
while data for $r=48$ and $r=64$ collapse, significant deviations appear for $r=192$.
The LRD becomes narrow and more symmetric as $r$ increases, as expected.

A LRD representative of the KPZ class is also shown in Figs. \ref{RDcrsos2dgeral}a,b.
It differs from the CRSOS curves at the peak in the linear plot
(Fig. \ref{RDcrsos2dgeral}a) and at the right tail in the log-linear one
(Fig. \ref{RDcrsos2dgeral}b).
However, this occurs because these data have good accuracy; instead, if fluctuations
of $10\%$ were present at the peaks and the tails could not be properly sampled
(e. g. in the case of a single small microscopy image), then the VLDS and KPZ curves
might be indistinguishable.
Thus, comparison of LRDs to distinguish universality classes
requires good accuracy in the data and independent checks of the distribution peaks
(in linear-linear plots) and tails (in log-linear plots).

\begin{figure}[!h]
\includegraphics[width=8.5cm]{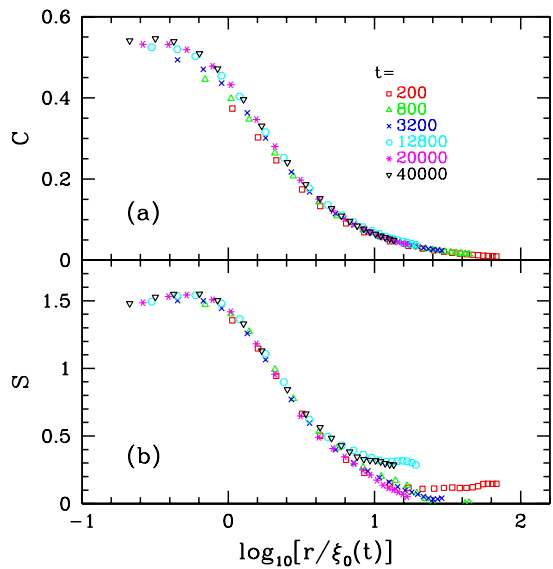}
\caption{(Color online) (a) Variation coefficient and (b) skewness of the LRDs of
the CRSOS model in $d=2$ as a function of the scaled box size at the listed times.
}
\label{ksscaledcrsos2d}
\end{figure}

\begin{figure}[!h]
\includegraphics[width=8.5cm]{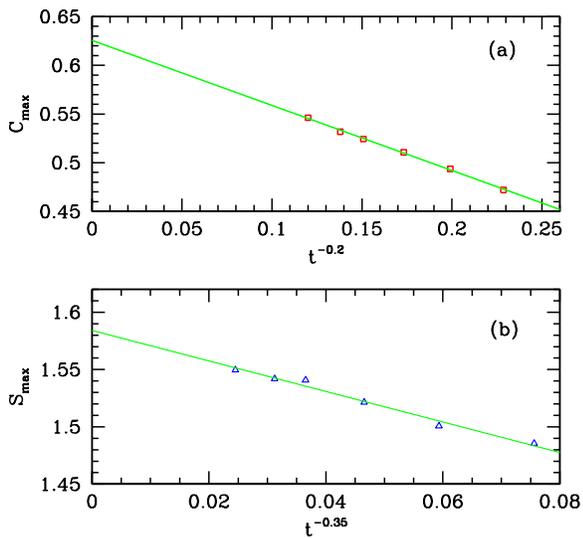}
\caption{(Color online) Extrapolation in time of the maximal variation coefficient (a)
and skewness (b) of the LRDs of the CRSOS model in $d=2$. The variables in the abscissa
have the exponents that give the best linear fits (solid lines) of each set of data.
}
\label{extrapcscrsos2d}
\end{figure}

Figs. \ref{ksscaledcrsos2d}a,b show $C$ and $S$ of the LRDs of the CRSOS model
as a function of $r/\xi_0$, confirming the scaling picture of Sec. \ref{scaling}.
These plots do not show plateaus of those quantities, but maximal values 
$C_{max}\left( t\right)$ and $S_{max}\left( t\right)$ may be extrapolated using
the same methods of Secs. \ref{simulation1d} and \ref{kpzsection}.
Figs. \ref{extrapcscrsos2d}a and \ref{extrapcscrsos2d}b show those quantities as
a function of $t^{-\lambda_C}$ and $t^{-\lambda_S}$, respectively, using
$\lambda_C=0.2$ and $\lambda_S=0.35$, which give the best linear fits of each data set
for $t\geq 1600$.
Asymptotic estimates are $C=0.63\pm 0.06$ and $S=1.58\pm 0.02$.

These values are larger than those obtained in the longest simulated time
($t=40000$), which are $C=0.54$ and $S=1.55$.
The larger discrepancy in the estimate of $C$ is related to the stronger time
dependence or, equivalently, to the smaller correction exponent $\lambda_C$.
For this reason, using the LRD at $t=40000$ and $r<100$ as representative
of the VLDS class may be misleading; for instance, the comparison with the KPZ curve
in Fig. \ref{RDcrsos2dgeral}a shows that they have approximately the same width ($C$),
although asymptotic estimates of the two classes differ almost $20\%$.

On the other hand, the skewness of the LRD of the CRSOS model has smaller scaling
corrections, which allows us to obtain an asymptotic estimate more accurate than $C$.
This skewness is much smaller than that of the KPZ class.

\section{Conclusion}
\label{conclusion}

LRDs in the growth regimes were calcutated in lattice models in the KPZ and VLDS classes
in $1+1$ and $2+1$ dimensions.
The emphasis on the KPZ LRDs is justified by its theoretical relevance and
recent experimental applications.

The LRD is expected to have a universal shape (stationary distribution) if calculated in
a range of box size $r$ that satisfies the relation $1\ll r\ll \xi\left( t\right)$,
where $\xi\left( t\right)$ is the correlation length, as proposed in previous works.
Here, plateaus of the dimensionless ratios $C$ (coefficient of variance) and $S$ (skewness)
are observed up to $r\approx 0.3\xi_1$, with $\xi_1$ defined as
the distance in which the autocorrelation function decreases to $10\%$ of its initial value.
For $r\gg \xi\left( t\right)$, we argue that the LRD converges to a Dirac delta function.
This leads to a universal crossover of those ratios as a function of $r/\xi$, which
is confirmed numerically.

The plateaus of $C$ and $S$ are narrow have non-negligible time dependence, even for
models that typically have small corrections in the average roughness scaling.
Thus, for a reliable quantitative characterization of the universal LRDs, an
extrapolation of the maximal values of those ratios is proposed and used to determine
their asymptotic values.
The consistency of the procedure is confirmed by results for the RSOS model in $1+1$
dimensions, in which the asymptotic estimates of $C$ and $S$ agree with those of
steady state EW interfaces in window boundary conditions \cite{antalpre}.
These asymptotic values may differ more than $10\%$ from the plateau values in the
maximal simulated times ($\sim {10}^4$ layers grown).
This shows that accounting for scaling corrections is essential to determine the
universality class of a given system by comparison of LRDs.

These results also confirmed the inadequacy of extrapolations of quantities such
as $C$ and $S$ to $r\to\infty$, since $r$ may eventually exceed $\xi$ and the LRD
will deviate from the universal shape.
This was formerly emphasized in Refs. \protect\cite{hhpal,hhtake}.
Moreover, our results show that scaling by the variance [Eq. (\ref{scalingvariance})]
should be avoided because it hinders the changes in $C$.
This is usually the most accurate quantity to characterize a distribution calculated
numerically, but it was not analyzed in former works on LRDs in the
growth regime.

In the CRSOS model, correlations propagate very slowly, thus plateaus of $C$ and $S$
are not observed even at long times.
Maximal values were extrapolated in time and provided estimates for the universal LRD.
Scaled LRDs of KPZ and VLDS models are visually similar in the usual log-linear
scales in two decades, which may turn difficult their use for comparisons with
low accuracy data.
This reinforces the relevance of the extrapolation procedures.

Comparison of RDs are generally believed to be advantageous over the comparison of
height distributions due to the typically small scaling corrections of the
steady state distributions.
However, the present results show that LRDs in the steady state regime may have
large scaling corrections and, consequently, their comparison may not be viewed as
a procedure superior to other approaches (e. g. calculation of scaling exponents or
height distributions), but as an important additional tool.

\vskip 1cm

{\bf Acknowledgements}

The author thanks Luis V\'azquez for helpful discussion and suggestions
to improve the method for comparing local roughness distributions.

This work was supported by CNPq and FAPERJ (Brazilian agencies).

%~~~~~~~~~~~~~~~~~~~~~~~~~~~~~~~~~~~~~~~~~~~~~~~~~~~~~~~~~~~~~~~~~~~~~~~~~~~
%~~~~~~~~~~~~~~~~~~~  REFERENCES  ~~~~~~~~~~~~~~~~~~~~~~~~~~~~~~~~~~~~~~~~~~
%~~~~~~~~~~~~~~~~~~~~~~~~~~~~~~~~~~~~~~~~~~~~~~~~~~~~~~~~~~~~~~~~~~~~~~~~~~~

\end{document}